\documentclass[aps,prb,onecolumn,preprint,amsmath,letter,showpacs,preprintnumbers,citeautoscript]{revtex4-1} 

\usepackage[french, english]{babel}  
\usepackage{amsmath} 
\usepackage{graphicx}
\usepackage{dcolumn}
\usepackage{bm}
\usepackage{subfigure} 
\usepackage{xspace} 
\usepackage{epstopdf}
\usepackage{multirow}

\newcommand{\ItoII}{$1 \rightarrow 2$ }
\newcommand{\IItoI}{$2 \rightarrow 1$ }
\newcommand{\ItoZ}{$1 \rightarrow 0$ }

\begin{document}


\title{Carrier recombination dynamics in InGaN/GaN multiple quantum wells}%

\author{Colin-N. Brosseau}
\author{Mathieu Perrin}
\altaffiliation[Present address: ]{Université Europ\'eenne de Bretagne, France, 
INSA, FOTON, UMR 6082, F-35708 Rennes.}
\author{Carlos Silva}
\author{Richard Leonelli}
 \email{richard.leonelli@umontreal.ca}
\affiliation{%
D\'epartement de Physique and Regroupement Qu\'eb\'ecois sur les Mat\'eriaux de Pointe, Universit\'e de Montr\'eal, Case Postale 6128, Succursale Centre-ville, Montr\'eal, Qu\'ebec H3C 3J7, Canada\\
}%

\date{\today}

\begin{abstract}
We have mesured the carrier recombination dynamics in InGaN/GaN multiple quantum wells over an unprecedented range in intensity and time by means of time-resolved photoluminescence spectroscopy.
We find that, at times shorter than 30\,ns, they follow an exponential form, and a power law at times longer than 1\,$\mu$s.
To explain these biphasic dynamics, we propose a simple three-level model where a charge-separated state interplays with the radiative state through charge transfer following a tunneling mechanism.
We show how the distribution of distances in charge-separated states controls the dynamics at long time.
Our results imply that charge recombination happens on nearly-isolated clusters of localization centers.
\end{abstract}

\pacs{78.47.-p, 78.47.Cd, 78.55.Cr, 78.67.De, }

\maketitle

\section{\label{sec:introduction}Introduction}

In$_x$Ga$_{1-x}$N/GaN heterostructures are essential for the production of optoelectronic devices, such as blue/ultraviolet lasers and light-emitting diodes, with external quantum efficiencies approaching 75\%.~\cite{nakamura00, gil02, narukawa:L963} 
The physical mechanisms that result in such high efficiencies are still poorly understood, as high threading dislocation densities due to epitaxial growth on mismatched sapphire substrates are belived to induce numerous non-radiative carrier recombination centers in these heterostructures.~\cite{rosner:420,kaneta:125317,sugahara:L398} 
It is generally assumed that, because of the large electron and hole effective masses in \mbox{InGaN}, nanoscopic potential fluctuations induce a strong localization of the carrier wavefunction.~\cite{schomig:106802,graham:103508} 
Evidence for carrier localization in In-rich clusters\cite{krestnikov:155310,bartel:1946} and antilocalization near threading dislocations\cite{hangleiter:2041,kaneta:125317,narukawa:981} has been reported.
Such a localization has a profound impact on carrier dynamics, which can be unraveled through time-resolved photoluminescence (PL) measurements. 
The PL decay from \mbox{InGaN/GaN} heterostructures is highly non-exponential~\cite{krestnikov:155310,bartel:1946,sun:49,pophristic:1114,chichibu:2051,onuma:151918,kwon:063509,lefebvre:35307} but a detailed understanding of these particular dynamics is still lacking.

Early studies  concluded that the recombination dynamics could be well described by a stretched exponential.~\cite{krestnikov:155310,pophristic:1114,chichibu:2051,bartel:1946,onuma:151918,kwon:063509}  
Stretched exponentials are often observed in highly disordered materials,\cite{sher:24} where they are assigned to the summation of monoexponential decays originating from individual localization centers with different exciton lifetimes.~\cite{bartel:1946} 
The quantum-confined Stark effet has also been invoked to explain the carrier dynamics.
It is induced in \mbox{InGaN/GaN} heterostructures by the large electric fields that result from the mechanical constraints inherently present in epitaxially grown nitrides, combined with high piezoelectric coefficients.
It leads to non-exponential recombination dynamics, redshift of the emission energy with time, and excitation density dependence on the emission energy.~\cite{lefebvre:35307}
The quantum-confined Stark effet is mostly observed in wide \mbox{InGaN} quantum wells (QW), since in QWs whose thickness is less than the exciton Bohr radius the \textit{e-h} overlap integral is not modified by the presence of electric fields.~\cite{bigenwald:371} 
The localization of electrons and holes on separate sites has also been reported.~\cite{morel:45331,kalliakos:428,hangleiter:2041} 
The amphoteric character of the localization centers has led to proposals of a bidimensional donor-acceptor like (2D-DAP) recombination model where electrons and holes are localized on the opposite sides of the QWs. 
The 2D-DAP model was found to reproduce the PL decay dynamics of \mbox{InGaN/GaN} QWs over three decades in intensity.~\cite{morel:45331}
Recently, the non-exponential PL dynamics of ensembles of localized centers have been linked to the PL intermittency, also called blinking, of individual centers whose emission is observed under continuous excitation conditions.~\cite{sher:101111,dunn:035330} 
Blinking is observed in a variety of systems such as colloidal and self-assembled quantum dots (QDs)\cite{nirmal:802,sher:101111,kuno:557,bertram:2666} and Si nanocristals,\cite{dunn:035330} and is attributed to the formation of  metastable, charge-separated (CS) dark states.\cite{tachiya:081104,zhao:157403}

In this paper, we present ensemble measurements of the carrier recombination dynamics in \mbox{InGaN/GaN} QWs over a time window that allows us to follow the PL decay over six decades in intensity. 
We observe a previously unreported slowing down of the decay dynamics at times longer than 200\,ns that cannot be explained by  stretched exponential, quantum-confined Stark effet or 2D-DAP models. 
We propose a  three-level recombination model involving  a dark CS state to explain our results.
We conclude that, in thin InGaN/GaN QWs, radiative recombination happens on nearly-isolated islands of agglomerated localization centers.

\section{Experimental details}

The \mbox{InGaN/GaN} samples were grown on c-plane sapphire substrates by metalorganic chemical vapor deposition.
Five-period In$_x$Ga$_{1-x}$N/GaN QWs were grown on 2 $\mu$m n-GaN with a nominal In composition $x=0.2$. 
The thickness of the GaN barriers  was  14.0 and that of the \mbox{InGaN} wells between 2 and 4 nm.
On top of the multiple QW structure, 14 nm of p-AlGaN and 100 nm of p-GaN were deposited.

Time-resolved PL measurements were performed using a  Ti:sapphire laser system with a repetition rate of 1 kHz and a pulse duration of 40\,fs. 
The second harmonic (390\,nm) of the laser beam, generated with a type-I $\beta$-BBO crystal, was focussed on the samples with an excitation density of about $3\times 10^2$\,W\,cm$^{-2}$, corresponding to $6 \times 10^{12}$\,photons\,cm$^{-2}$ per pulse. 
This excitation density is well within the linear regime of our samples.
The samples were cooled to 9\,K in a cold-finger  liquid-helium continuous-flow cryostat.
As the excitation photon energy (3.18\,eV) is lower than the low-temperature GaN band gap (3.5\,eV), the carriers were photogenerated directly in the QWs. 
The PL signal was analyzed with a 0.3-m grating spectrometer and detected with a gated, intensified charge-coupled device.
The time evolution of the emission spectrum was acquired with a temporal resolution of 5\,ns.

\section{\label{sec:results}Results and discussion}


\begin{figure}[b]
  \centering
  \includegraphics{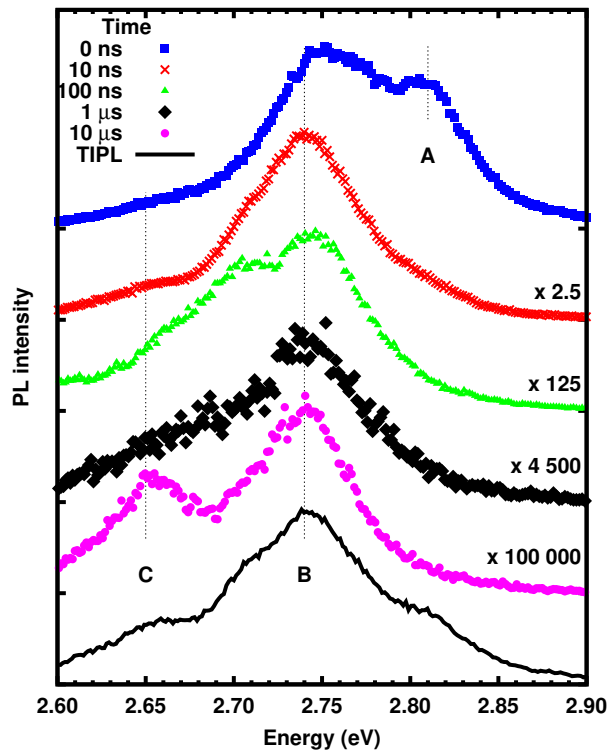}
  \caption{Time-resolved PL spectra of the sample with 2 nm QWs. TIPL is the time-integrated emission. $E_{exc} = 3.18$\,eV.}
  \label{fig:PL-t}
\end{figure}

The PL spectra of our samples showed multimodal emission spanning the range 2.2--3.1 eV. All samples exhibited  non-exponential decay dynamics with similar temporal dynamics. 
The time-resolved PL spectra of the sample with 2 nm QWs are presented in Fig.~\ref{fig:PL-t}.
Three bands can be distinguished.
Band A, at 2.81\,eV, has a lifetime shorter than 4\,ns. 
We attribute it to QW free-exciton recombination.
Band B peaks at 2.74\,eV. 
After a rapid  energy shift for $t<4$\,ns, its shape and energy position  remain constant  with time. 
At times longer than 1\,$\mu$s, band C can be observed at 2.65\,eV. 
The time-integrated PL spectrum is close to that of the time-resolved emission at 10\,ns. 
As band B dominates the emission, we focus on its dynamics in what follows.


\begin{figure}[h!]
  \centering
  \subfigure{\label{fig:Publi-mars09-plrt-semilogy-inset} \includegraphics{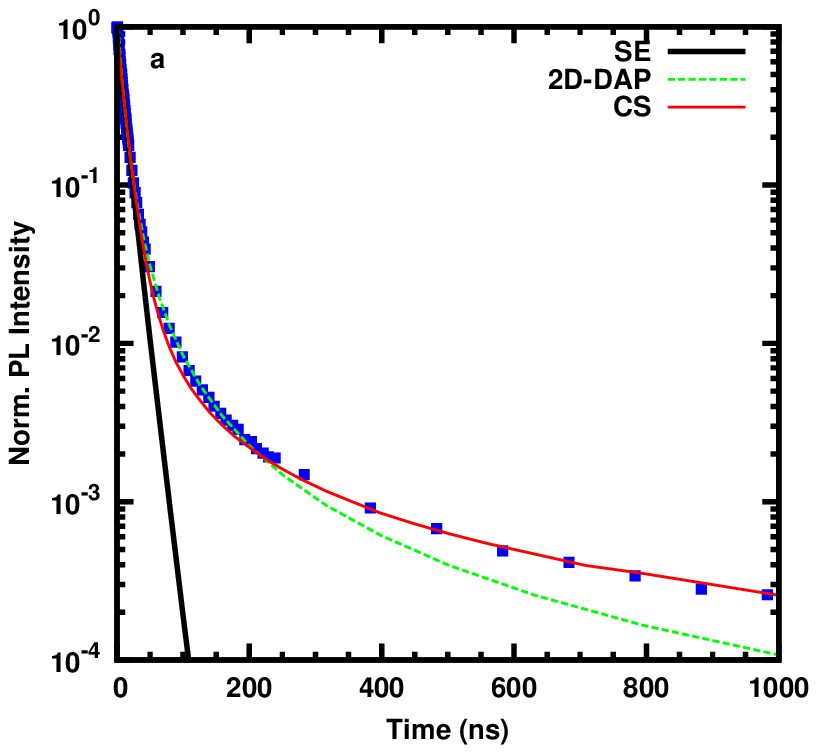}}
  \subfigure{\label{fig:Publi-mars09-plrt-loglog} \includegraphics{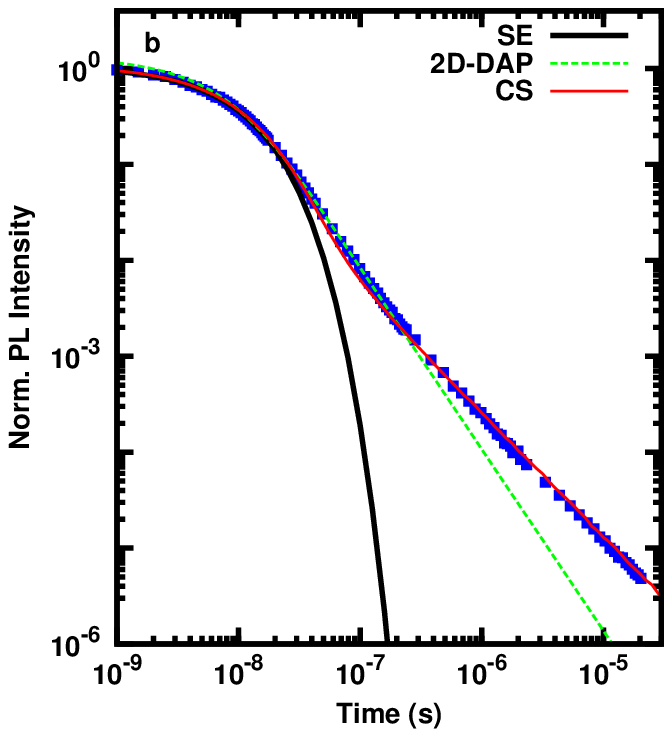}}
  \caption{ 
PL decay of band B. a) semilogarithmic plot; b) double logarithmic plot.
The solid lines are the best fits using the stretched exponential (SE), 2D-DAP, and CS models.}
  \label{fig:Publi-mars09-plrt}
\end{figure}

Figure\,\ref{fig:Publi-mars09-plrt} shows the decay of the emission from band B integrated from 2.73 to 2.76\,eV. 
For  times shorter than 30\,ns, the intensity decreases rapidly and its dynamics are compatible with a stretched exponential. 
At longer times however, the dynamics slow down and tend to a power law $t^{-\alpha}$ with $\alpha=1.3$ for $t>200$\,ns, as can be seen on Fig.\,\ref{fig:Publi-mars09-plrt-loglog}.
The long-lived tail on the decay curve, which accounts for about 20\% of the time integrated intensity, imposes stringent constraints on any  dynamical model.

The absence of spectral shift with time, the invariance of the decay curve with excitation density, together with the small thickness of the \mbox{InGaN} QWs, rules out the presence of a quantum-confined Stark effect in our samples.
Further, the significant proportion of photons emited in the long-lived tail implies that the power law it is not caused by a rare occurence of localization centers or impurities.
In addition, the shape preservation of bands with time and the energy independence of the decay dynamics, indicates the presence of a single recombination mechanism.

Within the stretched exponential model, the PL intensity is assumed to follow
\begin{equation}
\label{eq:SE}
I(t)=I_0 \exp\left\{-\left(\Gamma t \right)^\beta\right\}.
\end{equation}
As can be seen in Fig.~\ref{fig:Publi-mars09-plrt-semilogy-inset}, Eq.~\ref{eq:SE} reproduces our data with the parameters listed in Table \ref{tab:paramfits} only for $t<30$\,ns, that is for less than two decades in intensity.  
Thus, in our samples, the stretched exponential model is at best a phenomenological description at short times.

The 2D-DAP model implies a decay dynamics that follow\cite{morel:45331}
\begin{align}
  \label{eq:dap_2D_I_t}
  I(t) =&  2\pi n \int_0^\infty W(r) e^{-W(r)t} r \mbox{d}r   \nonumber \\
        & \times   \exp \left\{ 2\pi n \int_0^\infty \left( e^{-W(r)t} -1 \right) r \mbox{d}r \right\}, 
\end{align}
where $W(r)=\Gamma \mbox{e}^{-r^2/a^2}$ represents the recombination rate for an electron-hole pair separated by a distance $r$, $a$ is the Bohr radius, and $n$ is the majority carrier saturation density. Eq. \ref{eq:dap_2D_I_t} implies that the temporal shape of the PL decay remains the same in all samples where $\gamma=na^2$ is constant. Such a time-scale invariance was indeed reported by Morel~\textit{et al.}\cite{morel:45331}

Figure \ref{fig:Publi-mars09-plrt}  shows the best fit of our data with Eq.~\ref{eq:dap_2D_I_t}. 
A good agreement  is achieved for times up to about 200\,ns with the optimized parameters listed in Table~\ref{tab:paramfits}. The value obtained for $\gamma$ in our samples is close to that obtained by Morel~\textit{et al.}\cite{morel:45331} This indicates that our sample has a similar short-time decay dynamics.  
However, the 2D-DAP model fails to reproduce the decay slowing down observed at $t>200$\,ns.

\begin{table}[t]
\caption{Parameters obtained by fitting the stretched exponential (SE), 2D-DAP ans CS models to the experimental data. The numbers in parenthesis are the uncertainties on the values of the parameters.}
\label{tab:paramfits}
  \begin{ruledtabular}
    \begin{tabular}{lrcccc}
      
      && $\Gamma$ & $\beta$ & $\gamma$ & $\mu$ \\
      && ($10^8$ s$^{-1}$)\\
      \hline
      SE & &1.22 (0.08) & 0.83 (0.09) & - & - \\
      2D-DAP & &1.1 (0.1) & - & 0.28 (0.02) & - \\
      \multirow{2}{*}{CS} &  $\Gamma_{rad} $: &0.84 (0.06)\footnotemark[1] & - & - & 0.30 (0.01) \\
      & $\Gamma_{esc}$: &11 (2)\footnotemark[1] \\
        \end{tabular}
    \end{ruledtabular}
    \footnotetext[1]{Values obtained with $\nu = 2\times 10^{13}$\,s$^{-1}$.}

\end{table}

\begin{figure}[t]
  \centering
  \includegraphics[width=0.8 \linewidth]{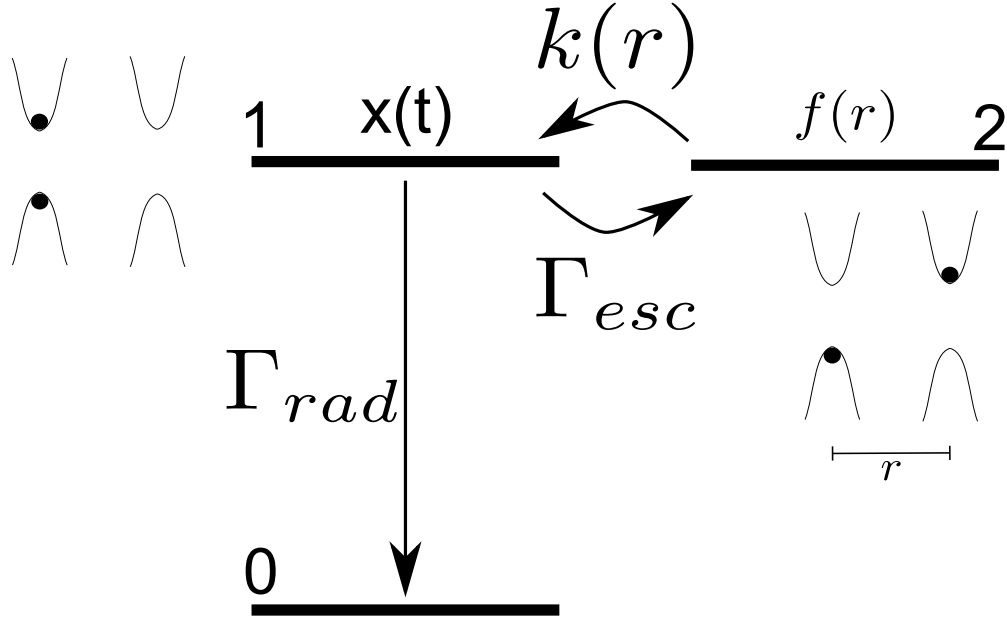}
  \caption{CS recombination model. 
State 1 corresponds to an electron and a hole  on the same localization center. 
State 2 corresponds to carriers  localized on different centers.  
Radiative transitions from state 1 to the ground state occur with rate constant $\Gamma_{rad}$ and charge separation from state 1 to state 2 occurs with a rate constant $\Gamma_{esc}$.}
  \label{fig:schema-modele-JMS}
\end{figure}

Sher~\textit{et al.} introduced a three-level CS model  to describe both PL blinking and non-exponential recombination dynamics in colloidal quantum dots.~\cite{sher:101111}
It is based on the interplay between a \textit{bright state} (1) that corresponds to an electron and a hole on the same localization center, and a \textit{dark state} (2) where the carriers are spatially separated, as schematized in Fig.~\ref{fig:schema-modele-JMS}.
Luminescence comes from  transitions \ItoZ with a rate constant  $\Gamma_{rad}$, while transitions 
 \ItoII  occur with a rate constant  $\Gamma_{esc}$. 
 Within the CS model, the charge separation induces a local modification of the potential landscape that affects the back-transfer \mbox{\IItoI.} If one assumes that the back-transfer occurs by tunnelling, the distribution of time spent in the dark state is
\begin{equation}
  R(t) = \int_0^{\infty}  f(r) k(r) e^{-k(r)t} \mbox{d}r,
\end{equation}
where $r$ is the distance between the carriers, $k(r) = \nu e^{-\eta r}$ is the tunnelling rate, and $f(r)$ is the spatial distribution of charge separation.~\cite{tachiya:081104}
Within the CS model, the PL intensity is given by
\begin{subequations}
  \begin{equation}
    I(t) \propto \Gamma_{rad}  x_1(t),
  \end{equation}
  where $x_1$ is the population of state 1 and
  \begin{align}
    \label{eq:dxdt}
    \dfrac{\mbox{d}x_1(t)}{\mbox{d}t} = & - \left( \Gamma_{rad} + \Gamma_{esc} \right) x_1(t) \nonumber \\
    & + \int_0^t  \Gamma_{esc} x_1(t') R(t-t') \mbox{d}t'.
  \end{align}
\end{subequations}
In the Laplace domain,
\begin{subequations}
  \begin{equation}
  \label{eq:xofs}
    \hat x_1(s) = \left( s + \Gamma_{rad} + \Gamma_{esc} [1-\hat R(s)] \right)^{-1}, 
  \end{equation}
  with
  \begin{equation}
   \hat{R}(s) =  \int_0^\infty \dfrac{f(r)k(r)}{s+k(r)} dr. 
   \end{equation}
   \end{subequations}  
If an exponential spatial distribution is assumed, $f(r) = \epsilon e^{-\epsilon r}$ and
  \begin{align}
    \label{eq:Rofs}
    \hat{R}(s) = \mu \int_0^\infty \dfrac{e^{-(1+\mu)x}}{s/\nu + e^{-x}} dx,
  \end{align}
where $x=\eta r$ and $\mu=\epsilon/\eta$.
$\hat{R}(s)$ can be evaluated through  numerical integration of Eq.~\ref{eq:Rofs} and $I(t)$ by an inverse Laplace transform of Eq.~\ref{eq:xofs} using the Gaver-Stehfest algorithm.~\cite{stehfest:47}
The CS model predicts biphasic dynamics.
At short times, the PL decays exponentially, $I(t) \propto e^{-\Gamma_{rad}t}$, while it follows a power law at long times, $I(t) \approx t^{-(1+\mu)}$.
Equations \ref{eq:dxdt} and \ref{eq:Rofs} indicate that $\nu$ and $\Gamma_{esc}$ cannot be determined independently.
For a given tunnel attempt frequency $\nu$,  $\Gamma_{esc}$ controls the strength of the long-time power law component with respect to the short-time exponential one.

As shown in Fig. \ref{fig:Publi-mars09-plrt}, the CS model reproduces the experimental results throughout the entire range of the measurement with a vibronic-like attempt frequency\cite{morteani:1708} and the optimized parameters listed in Table \ref{tab:paramfits}.
However, excellent agreement can still be obtained over a wide range of attempt frequencies; our simulations show that $\Gamma_{esc} \gtrsim \Gamma_{rad}$ and $\nu > 10^{10}$\,s$^{-1}$.

Within our model, the distribution of distances $f(r)$ controls the recombination dynamics.
We have tested that only rapidly decreasing functions, such as an exponential, can reproduce our results.
This implies that carriers are confined in nearly isolated islands of grouped localization centers.
Thus, one can hypothesize that many localization centers are in fact present in a single quantum dot.

The CS model is based on microscopic physical processes similar to the 2D-DAP model, as both models involve charge separation.  Within the CS model however, the charge separated state is dark and can be recycled from the bright state. It is this recycling that causes the decay slowing down observed at long times. As indicated above, the 2D-DAP model explains nicely the PL decay shape invariance reported in similar InGaN/GaN samples. The CS model also obeys a time-scale invariance law. This can be seen from Eq.~\ref{eq:xofs}, which can be approximated by~\cite{tachiya:081104} 
\begin{equation}
  \label{eq:xofsapp}
    \hat x_1(s) \approx \left[ s + \Gamma_{rad} + \Gamma_{esc} \dfrac{\pi \mu}{\sin(\pi\mu)}\left(\dfrac{s}{\nu}\right)^\mu \right]^{-1}.
  \end{equation}
Eq.~\ref{eq:xofsapp} breaks down for $\mu=1$ but reproduces well the decay dynamics for $\mu<1$. It follows  that the temporal shape of the PL decay is invariant if $\Gamma_{rad}^{\mu-1}\Gamma_{esc}\nu^{-\mu}$ remains constant.  Thus, within the CS model, the radiative rate of the bright state $\Gamma_{rad}$ is linked to the escape rate to the dark state $\Gamma_{esc}$ and to the back-transfer attempt frequency $\nu$. Further experimental and theoretical work is required to ascertain the physical origin of this time-scaling law.

Our results also allow to estimate the energy  barrier between the \textit{dark} and \textit{brigth state}, $E_{loc}$, from the back-transfer tunnel rate $\eta = \left( 8 m E_{loc} \hbar^{-2} \right)^{1/2} $.
Given the composition of our wells, one can estimate the electronic effective mass as $m = 0.18$\,m$_0$.~\cite{rinke:075202}
The fit of recombination dynamics gives $\mu = 0.30$.
Assuming intra-dot carrier localization, the characteristic distance between localization centers is then $1/\epsilon \approx$\,3-6\,nm.~\cite{narukawa:981,chichibu:4188,soshnikov:621}
One then estimates $E_{loc} \approx$\,25-50\,meV.

The proposed model for recombination dynamics in \mbox{InGaN} also seems to be  valid in a variety of materials such as colloidal QDs~\cite{sher:101111} or polymer distributed heterostructures.~\cite{gelinas:2010}
This indicates that the interaction of a charge-separated state with a radiative state plays a crucial role in the determination of long-time carriers recombination dynamics in systems with some degree of disorder.

\section{Conclusion}

We have measured the time-resolved photoluminescence of \mbox{InGaN/GaN} multiple quantum wells over an unprecedented range in intensity.
The dynamics show a slowing down at long times that cannot be explained by currents models.
Rather, it can be well modelled by a simple three-level system with a \textit{dark}  charge-separated state  and a \textit{bright} radiative state.
The interplay between the \textit{dark} and \textit{bright} states, throught back-transfer tunnelling, dominates the recombination dynamics at long times.
Our results indicate that the distribution of distance between charge-localization centers decreases rapidly with distance.
This implies that the radiative recombination happens on nearly-isolated islands of agglomerated localizations centers.

\begin{acknowledgments}
This work was supported by the Natural Science and Engineering Research Council of Canada (NSERC) and by the Fond Qu\'eb\'ecois de la Recherche sur la Nature et les Technologies (FQRNT). 
CS also acknowledges financial support from the Canada Research Chair Program.
\end{acknowledgments}


\end{document}